\documentclass{ws-rv961x669}
\usepackage{ws-rv-van}     % numbered citation/references (default)
\usepackage{ws-rv-thm}     % comment this line when `amsthm / theorem / ntheorem` package is used
\usepackage{subfigure}     % required only when side-by-side / subfigures are used
\makeindex
\usepackage{tabularx}

\begin{document}

\chapter[Universal scaling function ansatz for finite temperature jamming]{Universal scaling function ansatz for finite temperature jamming\label{LiuChapter}}
\author[S.A. Ridout, A. J. Liu and J.P. Sethna]{Sean A. Ridout, Andrea J. Liu and James P. Sethna}

\address{Department of Physics and Astronomy, University of Pennsylvania, Philadelphia, PA 19104 USA\\Department of Physics, Emory University, Atlanta, GA 30322 USA\\
LASSP, Cornell University, Ithaca, NY 14850 USA}

\begin{abstract}
We cast a nonzero-temperature analysis of the jamming transition~\cite{Degiuli2015} into
the framework of a scaling ansatz. We show that four distinct regimes for scaling exponents of thermodynamic derivatives of the free energy such as pressure, bulk and shear moduli, can be consolidated by introducing a universal scaling function with two branches. Both the original analysis and the scaling theory assume that the system always resides in a single basis in the energy landscape. The two branches are separated by a line $T^*(\Delta \phi)$ in the $T-\Delta \phi$ plane, where $\Delta \phi=\phi-\phi_c^\Lambda$ is the deviation of the packing fraction from its critical, jamming value, $\phi_c^\Lambda$, for that basin. The branch for $T<T^*(\Delta \phi)$ reduces at $T=0$ to an earlier scaling ansatz~\cite{Goodrich2016} that is restricted to $T=0$, $\Delta \phi \ge 0$, while the branch for $T>T^*(\Delta \phi)$ reproduces exponents observed for thermal hard spheres. In contrast to the usual scenario for critical phenomena, the two branches are characterized by different exponents. We suggest that this unusual feature can be resolved by the existence of a dangerous irrelevant variable $u$, which can appear to modify exponents if the leading $u=0$ term is sufficiently small in the regime described by one of the two branches of the scaling function. 

\end{abstract}

Like crystallization, jamming provides a paradigm for how particulate systems become rigid. For ideal spheres that increasingly repel each other as they overlap and do not interact if they do not overlap, the onset of rigidity with increasing packing fraction at zero temperature $T$ is discontinuous for the most stable possible state, the perfect FCC crystal, and critical for the least stable possible state, the jammed state at the jamming transition at a packing fraction $\phi_c$~\cite{OHern:2003vq,Goodrich:2014iu}. While crystallization is a straightforward first-order transition, the extent to which the jamming transition can be described within the framework of critical phenomena is still unclear.
Certainly many quantities exhibit power-law scaling, but are those scalings consistent with universal scaling 
functions with associated scaling relations among exponents?

In contrast to normal critical transitions, in which a diverging correlation length corresponds to fluctuations that diverge at the critical point, the jamming transition displays diverging correlations of fluctuations that \emph{vanish} at the transition~\cite{Hexner2018}. Moreover, the jamming transition point
depends on the history of the material -- the protocol with which one 
increases the density and decreases the temperature to reach jamming.
It is encouraging, however, that a universal scaling
ansatz~\cite{Goodrich2016} successfully described the zero-temperature elastic properties. Moreover, a jamming-like transition in a diluted
spring network~\cite{LiarteLub2016} has been cast into a universal scaling
form (see Supplement~\cite{LiarteLub2016}), and since turned into a general linear
response theory~\cite{PhysRevE.106.L052601,ThorntonLSnn}.

The situation at nonzero temperature $\tilde T$ is more challenging because the systems
are evolving with time. Work by De Giuli, et al.~~\cite{Degiuli2015} has bypassed this difficulty by assuming that at $\tilde T>0$ the system always resides in a given basin in the energy landscape, whose jamming transition is at $\phi_c^\Lambda$. That is, they describe a system near jamming using an effective contact network with contacts between particles that collide. They use the
well-studied force distribution of jammed packings to add the forces
generated by temperature-induced collisions across the gaps, analogous to methods developed to study stresses
in experiments measuring colloidal particle positions~\cite{LinBSSC16}.
This description is valid on sufficiently short timescales where this contact network does not change, analogous to methods developed to study stresses in experiments measuring colloidal particle positions~\cite{LinBSSC16}. While this assumption must break down at long time scales, it is reasonable to ask whether the scaling ansatz can be extended to nonzero temperatures within this approximation. We note that the zero-temperature scaling ansatz makes a similar approximation, because the protocol and 
sample-to-sample fluctuations of the critical jamming packing fraction
$\phi_c$ are swept under the rug by constructing a scaling ansatz in
terms of $\Delta \phi=\phi-\phi_c^\Lambda$, where $\phi_c^\Lambda$ can 
vary among configurations $\{ \Lambda \}$.

The theoretical arguments of De Giuli, et al.~\cite{Degiuli2015} apply to systems of $N$ soft, frictionless spheres in a volume $V$ at temperature $\tilde T$. The packing fraction is $\phi = \frac 1V \sum_i V_i$, where $V_i$ is the $d$-dimensional volume of particle $i$. The spheres interact via pairwise harmonic repulsions when they overlap, with a spring stiffness $k$. We introduce the dimensionless free energy, pressure, shear stress, bulk modulus and shear modulus using $k$ with appropriate factors of $D_\text{avg}$, the average particle diameter, so that $F=\textrm{free energy}/k D_\text{avg}^2$, $p=\textrm{pressure}*D_\text{avg}/k$, etc..  Note that we are departing from the usual custom of using temperature $T$ to construct a dimensionless free energy; this is important because we specifically interested in describing behavior at $T=0$ as well as $T >0$. Instead, we introduce the dimensionless temperature $T=\tilde T/k D_\text{avg}^2$.   The $T=0$ scaling ansatz~\cite{Goodrich2016} is expressed in terms of these dimensionless quantities for the same reason.

\begin{figure} \label{figphase}
\includegraphics[width=\columnwidth]{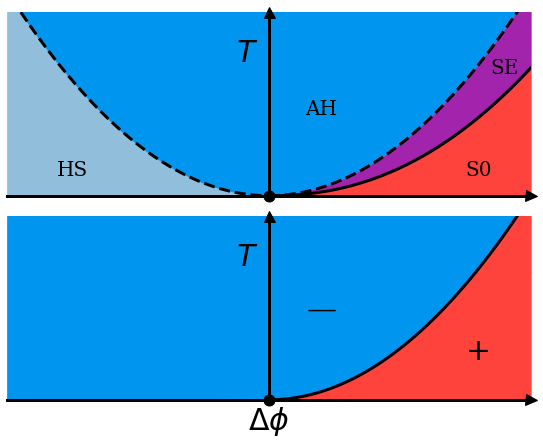}
\caption{
\label{fig:PhaseDiagram}
Top: The phase diagram of De Giuli et al\cite{Degiuli2015}.  Different colours, separated by lines, indicate four distinct scaling regimes, which are not necessarily separated by sharp phase transitions.  Bottom: The phase diagram corresponding to the scaling ansatz Equation \ref{eq:scaling_ansatz}. The $-$ branch reproduces the scaling exponents of three of De Giuli et al's scaling regimes. The phase
boundary lies along a constant ratio $T^*/\Delta\phi^{\zeta_+/\delta_{\phi,+}} = C^*$ in the notation of Eq.~\ref{eq:scaling_ansatz} (see footnote on page~\pageref{foot:PlusRegion}).}
\end{figure}

The marginal stability arguments and effective medium theory of De Giuli, et al.~\cite{Degiuli2015} suggest that there are four different scaling regimes in the $T-\phi$ plane~\cite{Degiuli2015},
Fig.~\ref{figphase}(top),
only one of which is described by the existing scaling ansatz~\cite{Goodrich2016}. The predicted exponents in the different regimes are consistent with mean field solutions for soft spheres~\cite{Zamponi2019} and the perceptron model~\cite{Franz2023}, although these later investigations showed only a crossover,
not a sharp transition -- perhaps the smearing expected by the evolving
landscape at finite temperatures. 

The existence of four distinct scaling regimes all terminating at the same critical point could suggest rather exotic criticality.  Here we show that the results of De Giuli, et al.~\cite{Degiuli2015} are captured by a scaling function in a manner similar to that of ordinary criticality, with two branches (Fig.~\ref{figphase}(bottom) separated by a continuous phase transition. This scaling function fully describe all of the scaling behaviors seen near the jamming transition in the full space described by the jamming phase diagram, namely temperature $T$, shear strain $\epsilon$, and packing fraction $\Delta \phi = \phi-\phi_c$, where $\phi_c$ is sample dependent. Unlike in ordinary criticality, the two branches are characterized by different exponents. We suggest that the difference in exponents can result from the existence of a dangerous irrelevant variable $u$, which can appear to modify exponents if the leading $u=0$ term is small in the regime described by one of the two branches of the scaling function.

Each state is also characterized by average number of interacting neighbors per particle (the contact number, $Z$), which satisfies $Z \ge Z_\textrm{min}$ where $Z_\textrm{min} = 2d - (2d-2)/N$ approaches the isostatic value $Z_\textrm{iso} \equiv 2d$ at the jamming transition in the thermodynamic limit~\cite{OHern:2003vq,Goodrich:2012ck}; we define $\Delta Z=Z-Z_\textrm{min}$.
The scaling ansatz of Ref.~\citen{Goodrich2016} was written for the elastic energy with the excess coordination beyond isostaticity, $\Delta Z$, as the control parameter analogous to reduced temperature $t$ in the Ising model. We will follow suit, constructing a scaling ansatz for the dimensionless free energy $F$ instead of the dimensionless elastic energy $E$ in order to describe behavior at $T \ge 0$:
%
%%\begin{widetext}
\begin{equation}
F\left(\Delta Z,\Delta \phi,\varepsilon,N,T\right) = \Delta Z^{\zeta_\pm} \mathcal{F}_\pm \left(\frac{\Delta \phi}{\Delta Z^{\delta_{\phi,\pm}}}, \frac{\varepsilon}{\Delta Z^{\delta_{\varepsilon,\pm}}},  N \Delta Z^{\psi_\pm} , \frac{T}{\Delta Z^{\zeta_\pm}} \right),
\label{eq:scaling_ansatz}
\end{equation}
%%\end{widetext}
%
where $\pm$ denote the two branches of the scaling function and the scaling exponents for the two branches, $\zeta_\pm$, $\delta_{\phi,\pm}$, $\delta_{\varepsilon,\pm}$ and $\psi_{\pm}$, are to be determined. Eq.~\ref{eq:scaling_ansatz} is set up so that the leading singular part of the free energy in the thermodynamic limit is proportional to $\Delta Z^\zeta_\pm$.  Note that $\zeta_\pm$ also appear in the temperature scaling; this follows simply from dimensional analysis~\cite{Goodrich2016}.

Note that the finite-size scaling is expressed in terms of the total number of particles, $N$, instead of the system length, $L \sim N^{1/d}$. We use $N$ instead of $L$ because of arguments~\cite{Goodrich:2012ck,Goodrich2016,hexner2019can,ikeda2020jamming} that the upper critical dimension for jamming is $d_u=2$.  Finite-size effects for $d \ge d_u$ should scale with $N$ with the exponent $\psi=d_u \nu$ where $\nu$ is the mean-field correlation length exponent~\cite{BINDER:1985vl}.

The excess packing fraction $\Delta \phi$ and shear strain $\varepsilon$ in Eq.~\ref{eq:scaling_ansatz} represent components of the same strain tensor (compression and shear respectively) but are allowed to scale differently.  Note that we assume systems are prepared isotropically so that different shear directions are statistically equivalent; this is why we represent them with a single $\varepsilon$. For systems prepared by shearing or by applying other loads, a more complicated formulation of the scaling ansatz is needed. For the $T=0$, $\phi>\phi_c$ case, Ref.~\citen{baityjesi2017} points the way towards such a formulation for shear-jammed systems. 

DeGiuli, et al.~\cite{Degiuli2015} identified four regimes in the $T-\phi$ plane  (Fig.~\ref{figphase}top): (1) The regime ``S0" in Ref.~\citen{Degiuli2015} corresponds to low $T$ at $\phi>\phi_c$. The exponents in this regime are those for $T=0$ soft spheres, described by the scaling ansatz of Ref.~\citen{Goodrich2016} (see Table~1 of Ref.~\citen{Degiuli2015} for a list of exponents in each regime). (2) A regime at somewhat higher temperatures at $\phi>\phi_c$, called ``SE" in Ref.~\citen{Degiuli2015}, or ``entropic soft spheres." In this regime the pressure is controlled by overlaps but the vibrational properties are strongly affected by thermal collisions.  (3) An anharmonic regime at packing fractions $\phi$ close to $\phi_c$ at nonzero $T$, called ``AH" in Ref.~\citen{Degiuli2015}. In this regime the pressure and other properties are purely entropic and controlled only by $T$. (4) A regime at $\phi<\phi_c$ and lower $T$, extending down to $T=0$, called ``HS" in Ref.~\citen{Degiuli2015} because it is described by the exponents near the jamming transition for thermal hard spheres. We will use the predicted exponents of De Giuli, et al.~\cite{Degiuli2015} in each of the four regimes to determine the exponents introduced in the scaling ansatz, and will show that they are fully consistent with the ansatz.

At $T=0$, $\phi>\phi_c$, Eq.~\ref{eq:scaling_ansatz} reduces to the scaling ansatz already proposed in Ref.~\citen{Goodrich2016}, with $\mathcal{F}_+=\mathcal{E}_0$ from Eq.~1 of Ref.~\citen{Goodrich2016}. Therefore, the exponents for the positive branch of the scaling function can be read off from Table 1 of Ref.~\citen{Goodrich2016}: $\zeta_+=4$, $\delta_{\phi,+} =2$, $\delta_{\varepsilon,+}=3/2$, $\psi_+=1$.  These exponents yield exponents for the pressure, bulk and shear moduli, etc. that are in agreement with those listed in the S0 regime in Ref.~\citen{Degiuli2015}; the exponents in this regime have been established for a long time~\cite{OHern:2003vq}. As in Ref.~\citen{Goodrich2016}, $\psi_+=d_u \nu_T$, where $\nu_T$ is the exponent for the transverse length scale, $\ell_T$. This length scale characterizes the wavelength of transverse phonons at the boson peak frequency, $\omega^*$~\cite{Silbert:2005vw,Liu:2010jx}: $\ell_T \sim c_T/\omega^*$ where the speed of transverse sound, $c_T$, scales as the square root of the shear modulus, $G$. Note that the boson peak frequency corresponds to the Ioffe-Regel crossover frequency where the phonon wavelength becomes comparable to the mean free path between phonon scattering events~\cite{Vitelli:2010fa}. Thus, $\ell_T$ also corresponds to the phonon mean free path at $\omega^*$.  Equivalently, $\ell_T$ corresponds to the length scale of correlations of polarization vectors in vibrational modes at $\omega^*$~\cite{Lerner:2013vf,Degiuli2015} (their $\ell_S(\omega^*)$). It is straightforward to show that $\ell_T  \sim \sqrt{G}/\omega^*$ has the same scaling as $\ell_S(\omega^*)$ in Table 1 of Ref.~\citen{Degiuli2015} in all four regimes S0, SE, AH and HS. 

The next step is to calculate the exponents for the negative branch of the scaling function. This can be done by comparing to exponents in the SE, AH or HS regimes of Ref.~\citen{Degiuli2015}. We will use the SE exponents, and then show that the resulting exponents for the negative branch are fully consistent with the exponents for the AH and HS regimes, as well, so that the negative branch describes all three regimes. 

We first consider the finite-size scaling behavior. The free energy must be continuous at all $T>0$ for any $\phi$ for any system size $N$. This implies that we must have equality of the finite-size scaling exponents for both branches: $\psi_-=\psi_+=1$. This implies a correlation length exponent of $\nu=\psi/d_u=1/2$, just as in the S0 regime. From Table 1 in Ref.~\citen{Degiuli2015}, we see that the length scale $\ell_T$ ($\ell_S(\omega^*)$ in their notation) scales as $(T/\Delta \phi)^{\frac{a-1}{4}}$ where $a=(1-\theta_e)/(3+\theta_e)$ and $\theta_e \approx 0.42$ is the exponent characterizing the low force tail of the force distribution in mean field theory~\cite{charbonneau2014exact, charbonneau2014fractal} , as well as in low dimensions when bucklers are excluded~\cite{Charbonneau2015}. Note that $\Delta Z$ scales as $(T/p)^{2b}$ where $p \sim \Delta \phi$ in this regime, where $b=(1+\theta_e)/(3+\theta_e)$. Thus, $\ell_T \sim \Delta Z^{-1/2}$, so that $\nu_T=1/2=\nu$. Thus, the relevant length scale controlling finite-size effects in both the S0 and SE regimes is the transverse length scale $\ell_T$.  

The remaining exponents can be obtained by differentiating Eq.~\ref{eq:scaling_ansatz} and comparing with the scalings of Table 1 of Ref.~\citen{Degiuli2015}. For example, by differentiating $F$ with respect to $\Delta \phi$, we find that the pressure satisfies $p \sim \Delta Z^{\zeta_- - \delta_{\phi,-}}$. But from Table 1, we see that $p \sim \Delta \phi \sim \Delta Z^{\delta_{\phi,-}}$, so $\zeta_-=2\delta_{\phi,-}$. But we also have $\Delta Z \sim \left(T/p\right)^{2b} \sim \Delta Z^{2 b\zeta_-}/\Delta Z^{2 b \left(\zeta_- - \delta_{\phi,-}\right)} \sim \Delta Z^{2b \delta_{\phi,-}}$. Thus, $\delta_{\phi,-}=1/(2b)=(3+\theta_e)/(1+\theta_e)$ and $\zeta_-=2(3+\theta_e)/(1+\theta_e)$. Similarly, $\delta_{\varepsilon,-}=(2+\theta_e)/(1+\theta_e)$.

It is straightforward to show that if one adopts the exponents from Table 1 of Ref.~\citen{Degiuli2015} in the hard- sphere (HS) regime, one obtains the same values of the exponents $\psi_-$, $\zeta_-$, $\delta_{\phi,-}$ and $\delta_{\varepsilon,-}$. Moreover, by adopting these exponents and applying the scaling ansatz in Eq.~\ref{eq:scaling_ansatz} at $\Delta Z=0$, $T>0$, where the behavior must be analytic, one recovers the exponents for the anharmonic (AH) regime of Table 1 of Ref.~\citen{Degiuli2015}. Thus, three of the four regimes, HS, AH and SE, of Fig.~1 in Ref.~\citen{Degiuli2015} (the blue and purple regions) all correspond to the negative branch of the scaling function with identical exponents (for details, see Appendix \ref{app:scaling_regimes}). The remaining regime (S0) is described by the positive branch of the scaling function. In every regime, the length scale controlling finite-size effects is the transverse length scale $\ell_T$ ($\ell_S(\omega^*)$ in the notation of Ref.~\citen{Degiuli2015}).

The free energy must be continuous everywhere, so the two branches of the scaling function must yield the same free energy along a curve $T^*(\Delta \phi,\varepsilon)$ for every system size $N$. Here we restrict ourselves to the case where $\varepsilon=0$ and $N \rightarrow \infty$ to compare directly to the prediction of Ref.~\citen{Degiuli2015}. Clearly we must have
%\begin{widetext}
\begin{equation}
 \Delta Z^{\zeta_+} \mathcal{F}^\infty_+ \left(\frac{\Delta \phi}{\Delta Z^{\delta_{\phi,+}}} , \frac{T^*}{\Delta Z^{\zeta_+}}\right)=\Delta Z^{\zeta_-} \mathcal{F}^\infty_- \left(\frac{\Delta \phi}{\Delta Z^{\delta_{\phi,-}}} ,\frac{ T^* }{\Delta Z^{\zeta_-}} \right). \label{eq:Tstar1}
 \end{equation}
 %\end{widetext}
Eliminating $\Delta Z$ from the equation (see Appendix \ref{app:eliminate_dz}), we obtain
\begin{equation}
 T^* \mathcal{\tilde F}^{\infty}_+ \left(\frac{\Delta \phi^{\zeta_+/\delta_{\phi,+}}}{T^*} \right)=T^* \mathcal{\tilde{F}}^\infty_- \left(\frac{\Delta \phi^{\zeta_-/\delta_{\phi,-}}}{T^*} , \right). \label{eq:Tstar}
 \end{equation}
Remarkably, we note that $\zeta_+/\delta_{\phi,+}=\zeta_-/\delta_{\phi,-}=2$. If these ratios were not equal for the two branches of the scaling function, continuity of $F$ would not be possible. Eq.~\ref{eq:Tstar} implies that $T^* \sim \Delta \phi^2$. This agrees with earlier predictions~\cite{Zhang:2066bv,Berthier2013,Goodrich2016} but differs from the results of Ref.~\citen{Degiuli2015}, who found $T^* \sim \Delta \phi^{(2-a)/(1-a)}$; our arguments reproduce their scalings without introducing this anomalous temperature scale.

The resulting phase diagram in the $T-\phi$-plane is as depicted in Fig.~\ref{figphase}. The exponents are summarized in Table~\ref{tab:exp}. The red regime is described by the positive branch of the scaling function while the blue regime is described by the negative branch. The curve $T^* \sim \Delta \phi$ describes the boundary between the two branches. The free energy is continuous at $T^*$ but its derivatives are not necessarily continuous in the limit $N \rightarrow \infty$, so $T^*(\Delta \phi)$ is a line of phase transitions. This can be checked by calculating derivatives of the functions $\tilde F_\pm$ for different $N$ and evaluating them at $T^*$. 

\begin{table}[ht]
%\begin{center}
\tbl{\label{tab:exp}Values of critical exponents for the two branches ($+$ and $-$) of the scaling ansatz of Eq.~\ref{eq:scaling_ansatz}.}
{\begin{tabular}{@{}ccc@{}}
\toprule
   & +  & - \\
\hline
  $\zeta$ & 4 & $2(3+\theta_e)/(1+\theta_e)$ \\
$\delta_\phi$ & 2 & $(3+\theta_e)/(1+\theta_e)$ \\
$\delta_\varepsilon$ & 3/2 & $(2+\theta_e)/(1+\theta_e)$ \\
$\psi$ & 1 & 1 \\
\botrule
\end{tabular}}
%\end{center}
\end{table}

For jammed soft spheres, Goodrich, et al.~\cite{Goodrich2016} argued for a scaling relation connecting shear stress to the pressure:
\begin{equation}
s^2 N/p^2=\mathcal{S_+}(Np^{1/2}).
\end{equation}
This holds for the positive branch of the scaling function. A consequence of the scaling ansatz is that for the negative branch there is anomalous scaling of shear stress fluctuations so that
\begin{equation}
s^2 N /p^x = \mathcal{S_-}(N p^{2b}), \label{eq:sminusscaling}
\end{equation}
at fixed $T/\Delta \phi^2$ or $T/p^2$, where $x=\frac{7+ \theta}{3+\theta} \approx 2.17$ and $2b=(1+\theta_e)/(3+\theta_e)$.  Because at fixed $p, T$ the correlation length is finite, the scaling function $\mathcal{S}_{-}{\left(X\right)}$ must approach a constant as $X\to \infty$.

As shown in Appendix~\ref{app:shear_stress}, for small $\left|\Delta \phi\right|$ the scaling ansatz predicts the anomalous scaling

\begin{equation}
   s^2 \propto \frac{T^{x/2}}{N} \approx \frac{T^{1.09}}{N}.
\end{equation}

This is a slightly stronger scaling than equipartition for a fixed spring network, consistent with the thermal stiffening of the effective contacts.

We note that the critical exponents in our scaling function differ on the two
sides of the transition. There is a longstanding tradition in statistical
mechanics of allowing for this possibility, but almost always the exponents
are the same -- as follows from the linearized flow of the renormalization group.
(The exponents are ratios of the eigenvalues of the flow at a fixed point,
and thus are equal on both sides of the fixed point.) Different sets of exponents for the two branches would seem to imply a highly unusual form of criticality.

It is possible, however, to generate this apparent
difference of exponents through an alternative scenario, which, though somewhat contrived, has the advantage of being consistent
with the traditional renormalization group picture.
Consider the effects of an irrelevant variable $u$, which we incorporate
by adding another argument $X_u = u \Delta Z^{\zeta_- - \zeta_+}$ to the new
scaling function $\mathcal{F}_u$.
Here, $u$ is irrelevant because $\zeta_- > \zeta_{+}$.  Since $u$ is
irrelevant, near the transition we may expand the scaling
function around $u=0$. We do so in an unorthodox way, evaluating the derivative
not at zero (as in a Taylor expansion) but at the current value
$X_u = u \Delta Z^{\zeta_- - \zeta_+}$:
%\begin{widetext}
\begin{equation}
\begin{aligned}
\label{eq:jim_scaling_ansatz}
F\left(\Delta Z,\Delta \phi,\varepsilon,T\right) 
=& \Delta Z^{\zeta_+} 
    \mathcal{F}_u\left(\frac{\Delta \phi}{\Delta Z^{\delta_{\phi,+}}}, \frac{\varepsilon}{\Delta Z^{\delta_{\varepsilon,+}}},  \frac{T}{\Delta Z^{\zeta_+}} , u \Delta Z^{\zeta_- - \zeta_+}\right).\\
 \approx& \Delta Z^{\zeta_+} 
	\left(\mathcal{F}_u\left(X^+_\phi , X^+_\varepsilon, X^+_T, 0\right)
    + X_u \mathcal{F}_u^{[0,0,0,1]}\left(X^+_\phi,X^+_\varepsilon,X^+_T,X_u\right)\right).
\end{aligned}
\end{equation}
where the invariant scaling combinations $X^{+/-}_x$ are those valid in
the plus/minus region. The first term 
$\mathcal{F}_u(X^+_\phi , X^+_\varepsilon, X^+_T, 0)$
equals $\mathcal{F}_+(X^+_\phi , X^+_\varepsilon, X^+_T)$, 
the traditional universal scaling function dominant as $\Delta Z \to 0$.
But suppose that this dominant term is zero in the minus region%
\footnote{\label{foot:PlusRegion}
The plus region is $X_\phi>0$, $X_T < C^* X_\phi^{\zeta/\delta_\phi} = C^* X_\phi^2$, see Fig.~\ref{fig:PhaseDiagram}.}
(as the 
scaling form for the Ising magnetization is zero for $T>T_c$). Then
the free energy in the minus region
will be given by the second term, subdominant by a factor $\Delta Z^{\zeta_- - \zeta_+}$, as desired. To get the scaling form of Eq.~\ref{eq:scaling_ansatz} for the 
minus region, this derivative must take the form (Appendix \ref{app:u_exponents})
\begin{equation}
\begin{aligned}
u \mathcal{F}_u^{[0,0,0,1]}\left(X^+_\phi,X^+_\varepsilon,X^+_T,X_u\right) 
&=\mathcal{F}_-\left(X^-_\phi,X^-_\varepsilon,X^-_T\right) \\
&=\mathcal{F}_-\left(u^{1/2} X^+_\phi/X_u^{1/2},u^{1/4} X^+_\varepsilon/X_u^{1/4},u X^+_T/X_u\right).
\end{aligned}
\end{equation}
for the particular value of $u$ for this system. This leads to
\begin{equation}
\begin{aligned}
F\left(\Delta Z,\Delta \phi,\varepsilon,T\right) 
 =& \Delta Z^{\zeta_+} 
	\mathcal{F}\left(\frac{\Delta \phi}{\Delta Z^{\delta_{\phi,+}}}, \frac{\varepsilon}{\Delta Z^{\delta_{\varepsilon,+}}},  \frac{T}{\Delta Z^{\zeta_+}}\right)\\
 &+ \Delta Z^{\zeta_-} 
     \mathcal{F}_-\left(\frac{\Delta \phi}{\Delta Z^{\delta_{\phi,-}}}, \frac{\varepsilon}{\Delta Z^{\delta_{\varepsilon,-}}},  \frac{T}{\Delta Z^{\zeta_-}}\right)
\end{aligned}
\end{equation}
%\end{widetext}
as desired.

One prediction of this picture is that it should be possible to find an irrelevant variable that, when tuned to a large value, induces corrections to the ordinary zero-$T$ soft sphere scaling that realize the hard-sphere exponents. Some changes to the potential 
%(e.g. increasing the exponent $\alpha$, as discussed above)
are irrelevant. Changes to the preparation protocol of states are also irrelevant, and it would be interesting to explore whether they could be responsible for these predicted corrections to scaling.

In summary, a scaling theory can be constructed for the jamming transition as a function of the variables encapsulated by the jamming phase diagram, namely packing fraction, shear strain and temperature. We find that the known behaviors are captured by two branches of a scaling function for the free energy. The two branches join together at a temperature $T^* \sim \Delta \phi^2$.

Recall that temperature $T$ is dimensionless: $T=\tilde T/k D_{\mathrm{avg}}^2$ where $\tilde T$ is the dimensional temperature. We can interpret the $+$-branch of the scaling function, corresponding to $T<T^*$, as the regime in which the behavior is dominated by the particle stiffness $k$ or the interaction energy. On the other hand, the $-$-branch, where $T>T^*$, is dominated by temperature, $\tilde T$, or thermal collisions.

We note that the arguments presented here can be generalized to a family of potentials described by the exponent $\alpha$:
\begin{equation}
U(r_{ij}) = \frac{U_0}{\alpha}\left(1-\frac{r_{ij}}{R_i+R_j}\right)^\alpha \Theta\left(1-\frac{r_{ij}}{R_i+R_j}\right), \label{eq:pairwise_interaction}	
\end{equation}
where $r_{ij}$ is the distance between the centers of particles $i$ and $j$ and $R_i$ and $R_j$ are the radii of the particles. Here $\Theta(x)$ is the Heaviside step function and $U_0$ is the interaction energy scale. Following earlier work~\cite{Vitelli:2010fa}, it is useful to define an effective spring constant, 
$k_{\textrm{eff}} =U_0 (\alpha-1)/D_\textrm{avg}^2 \Delta Z^{2(\alpha-2)}$, where $D_\textrm{avg}$ is the average particle diameter.  If we use $k_{\textrm{eff}}$ to construct dimensionless free energy, temperature, etc., as discussed before Equation \ref{eq:scaling_ansatz}, then the results are generally applicable for any $\alpha>1$.

The limit $\alpha=1$, where $k_{\mathrm{eff}} = 0$, is singular. There the de-dimensionalization of scaling variables using $k_{\textrm{eff}}$ clearly fails. In this case, scaling exponents are predicted to be zero and simulations and mean-field calculations by Sclocchi, Urbani, and Franz~\cite{Franz2019, Franz2020, Franz2021} show that logarithmic corrections are dominant. It would be interesting to see if the scaling ansatz applies to these logarithmic corrections as well.

In our formulation of the scaling ansatz, we used $\Delta Z$ as a control variable, following the approach of Goodrich, et al.~\cite{Goodrich2016}. This choice is appealing because it treats $\Delta \phi$ and $\varepsilon$--compressive and shear strain--in the same way. However, Goodrich, et al.~\cite{Goodrich2016} pointed out that we could alternatively choose $\Delta \phi$ as the control variable. This would lead to the scaling ansatz:
%\begin{widetext}
\begin{equation}
	F\left(\Delta Z,\Delta \phi,\varepsilon,N,T\right) = \Delta Z^{\zeta_\pm} \mathcal{F}_\pm \left(\frac{\Delta \phi}{\Delta Z^{\delta_{\phi,\pm}}}, \frac{\varepsilon}{|\Delta \phi|^{\delta_{\varepsilon,\pm}/\delta_{\phi,\pm}}},  N |\Delta \phi|^{\psi_\pm/\delta_{\phi,\pm}} , \frac{T}{ |\Delta \phi|^{\zeta_\pm/\delta_{\phi,\pm}}} \right), \label{eq:scaling_ansatz_phi}
\end{equation}
%\end{widetext}

Note, however, that this form requires using $|\Delta \phi|$ to scale $\epsilon$, $N$ and $T$, which causes problems because it introduces non-analyticity at $\Delta \phi=0$ even at $T>0$. It would seem, therefore, that $\Delta Z$ is a better choice for the control variable.
The discovery of a correlation length that characterizes $\Delta Z$ fluctuations and diverges at the jamming transition~\cite{Hexner2018} suggests still another choice for the control variable, namely the pressure, $p$, which is conjugate to $\Delta \phi$. So far, the length scale characterizing contact number fluctuations is the only diverging correlation length that has been identified for the jamming transition, although there are multiple diverging length scales have been identified that are not associated with correlation functions. It could be argued that $\ell_T$ is a diverging correlation length because it corresponds to the length scale for correlations in normal modes of vibrations at the boson peak frequency $\omega^*$. The length scale $\ell_S{\left(\omega^*\right)}$, however, involves the boson peak frequency $\omega^*$ and does not correspond to the correlation length of any of the variables that appear in the scaling ansatz. We note that the $\Delta Z$ correlation length (which is associated with $\Delta Z$, one of the variables in the scaling ansatz) does not appear to scale the same way as $\ell_S(\omega^*)$ (which controls finite-size scaling). This lack of correspondence between a correlation length and the length scale characterizing finite-size effects, while unusual, appears to be characteristic of other systems with sharp global transitions but configuration-dependent critical densities, as originally identified in the depinning of charge-density waves~\cite{MyersS93A,MyersS93B,Middleton:1993vc}.

While it is normal for a correlation length to diverge at a critical point, it is decidedly abnormal for the fluctuations to vanish at long length scales--usually they diverge. Contact number fluctuations must vanish at long length scales because the average contact number at the transition must be $Z=2d$, as dictated by isostaticity. As a result, contact number fluctuations are hyperuniform~\cite{Hexner2018, hexner2019can}, implying that there can be no square-gradient expansion--no Ginzburg-Landau theory--for jamming. It is not clear how to construct a renormalization group for a system that is hyperuniform at long length scales. The existence of other systems that exhibit hyperuniformity at transitions, such as systems with transitions that separate absorbing states from nonabsorbing ones~\cite{Hexner2015}, suggest that the challenge may not be restricted to the jamming transition. It is interesting and nontrivial that despite being highly unusual in exhibiting hyperuniformity, the jamming transition can nevertheless  be described by a scaling ansatz.

It should be noted, however, that the vanishing of $\Delta Z$ fluctuations may be specific to systems prepared at fixed pressure (pressure control). A recent study at fixed $\Delta \phi$ (phi-control) suggests that that $\Delta Z$ fluctuations might diverge instead of vanish and that even the finite-size scaling exponent is different in that case, at least for $\Delta Z$ and energy fluctuations~\cite{ikeda2023control}. It would be interesting to see whether a different scaling ansatz can be constructed for that case, and how the ansatzes for pressure- and $\phi$-control might be related.

Finally, we must again raise the caveat that the arguments of De Giuli, et al.~\cite{Degiuli2015} and our scaling ansatz are predicated on the approximation that the system is restricted to a single basin in the energy landscape. At $T>0$, we know that this is not true--given enough time, the system will explore multiple basins in the landscape, each with a minimum corresponding to a different value of the jamming critical packing fraction, $\phi_c$. It is likely that thermal fluctuations that drive the system over energy barriers will destroy the line of phase transitions at the boundary separating the two branches of the scaling function. In other words, glass physics may well smear the phase transition line into a crossover. However, the jamming transition and the line that emanates from it have such unusual properties in the canon of criticality, and are so important to our understanding of disordered solids, that it is important to understand them whether the line is singular or only demarcates a crossover.

\section*{Acknowledgements}

We thank G. Biroli, S. Franz, and F. Zamponi for useful discussions, the Simons Foundation for supporting this work through the "Cracking the glass problem" collaboration (\#454945, SAR and AJL), the NSF (DMR-1719490 to JPS), and the 2019 Beg Rohu Summer School, where this work was primarily done, as well as the Center for Computational Biology at the Flatiron Institute, where the manuscript was primarily written. This chapter is dedicated to the memory of Michael E. Fisher, the thesis advisor of Andrea J. Liu and colleague of J. P. Sethna. Liu, in particular, is grateful for having had the opportunity to absorb scaling theory and criticality directly from Fisher. We hope that this contribution helps to demonstrate the lasting value of looking at the world through the lens of critical phenomena.

\bibliographystyle{ws-rv-van}

\bibliography{bibliography}

\appendix

\section{Converting between $\Delta Z$ and $\Delta \phi$ in the scaling ansatz}
\label{app:eliminate_dz}
As in the zero-temperature case, \cite{Goodrich2016}, the scaling ansatz has the unusual feature that the variables in it are not all independent: at fixed $\Delta \phi, T, N$, and preparation protocol, the coordination $\Delta Z$ is fixed by some equation of state. 

In addition to the scaling ansatz, we assume a scaling equation of state 
\begin{equation}
 \frac{\Delta \phi}{\Delta Z^{\delta_{\phi,\pm}}} = f{\left(\frac{\varepsilon}{\Delta Z^{\delta_{\varepsilon,\pm}}},  N \Delta Z^{\psi_\pm} , \frac{T}{\Delta Z^{\zeta_\pm}} \right)}.
\end{equation}

If this is assumed, then it is possible to eliminate $\Delta Z$ from the scaling ansatz, in order to obtain Equation \ref{eq:Tstar} in the main text, or make comparisons to De Giuli et al's scaling predictions (which are expressed in terms of $\Delta \phi$ rather than $\Delta Z$).

This equation of state can be inverted to yield

\begin{equation}
	\Delta Z^{\delta_{\phi,\pm}} = \left| \Delta \phi \right| g_{\pm}{\left(\frac{\varepsilon}{\left| \Delta \phi\right|^{\delta_{\varepsilon,\pm}/ \delta_{\phi, \pm}}},  N \left| \Delta \phi \right|^{\psi_\pm / \delta_{\phi, \pm}} , \frac{T}{\left| \Delta \phi \right|^{\zeta_\pm / \delta_{\phi, \pm}}} \right)}.
\end{equation}

Here the branches $f_{\pm}$ correspond to the sign of $\Delta \phi$ rather than the regions $-, +$ of the phase diagram; if we wished to completely eliminate $\Delta Z$ from the scaling ansatz in all regions simultaneously it may introduce a nonanlyticity at $\Delta \phi = 0$.  However, this is not a problem if we only wish to eliminate $\Delta Z$ from an equation with the sign of $\Delta \phi$ fixed.

Subject to this restriction, this allows us to eliminate $\Delta Z$ from any equation of the original scaling variables:

\begin{align}
h{\left(\frac{\Delta \phi}{\Delta Z^{\delta_{\phi,\pm}}},\frac{\varepsilon}{\Delta Z^{\delta_{\varepsilon,\pm}}},  N \Delta Z^{\psi_\pm} , \frac{T}{\Delta Z^{\zeta_\pm}} \right)} &= \tilde{h}_{\pm}{\left(\frac{\varepsilon}{\left| \Delta \phi\right|^{\delta_{\varepsilon,\pm}/ \delta_{\phi, \pm}}},  N \left| \Delta \phi \right|^{\psi_\pm / \delta_{\phi, \pm}} , \frac{T}{\left| \Delta \phi \right|^{\zeta_\pm / \delta_{\phi, \pm}}} \right)}
\end{align}

Restricting to $\Delta \phi > 0$ we may thus eliminate $\Delta Z$ to obtain equation \ref{eq:Tstar} in the main text, and as necessary in the following section Appendix \ref{app:scaling_regimes} to verify that the scaling ansatz reproduces the scaling regimes of De Giuli et al.\cite{Degiuli2015}

\section{Deriving the De Giuli et al. exponents from the scaling ansatz}
\label{app:scaling_regimes}
Here we show in more detail how to obtain the various scalings described by De Giuli et al.~\cite{Degiuli2015} from the scaling ansatz.

We begin with the scaling behaviour of the pressure. Start with the ansatz of Equation (\ref{eq:scaling_ansatz}), and take $N\to\infty$ and $\epsilon = 0$.  Taking a derivative with respect to the first argument yields the scaling ansatz for the pressure, 

\begin{align}
P &=\Delta Z^{\zeta_{\pm} - \delta_{\phi,\pm}} \mathcal{P}_{\pm}{\left(\frac{\Delta \phi}{\Delta Z^\delta_{\phi, \pm}}, \frac{T}{\Delta Z^{\zeta_{\pm}}}\right)}.\end{align}

Eliminating $\Delta Z$ as above gives

\begin{align}
P &=\left| \Delta \phi\right|^{\zeta_{\pm}/\delta_{\phi, \pm} - 1} \mathcal{P}_{\pm}{\left(\frac{T}{\left|\Delta \phi\right|^{\zeta_{\pm}/\delta_{\phi, \pm}}}\right)}.\end{align}

First, we study the ``anharmonic'' limit $T \gg \Delta \phi^2$, in which De Giuli et al. predict $P \propto \sqrt{T}$.  We set $\Delta \phi \to 0$ in the scaling function at fixed $T$, and require that the behaviour of the scaling function is not singular in $\Delta \phi$. This requires that the asymptotic behaviour of the scaling function has the right power law to balance all powers of $\Delta \phi$, i.e.
\begin{align}
P &\propto \left|\Delta \phi\right|^{\zeta_{-}/\delta_{\phi, -}- 1} \left(\frac{T}{\left|\Delta \phi\right|^{\zeta_{-}/ \delta_{\phi,-}}}\right)^\frac{\zeta_{-} - \delta_{\phi, -}}{\zeta_{-}} \\ &= T^\frac{\zeta_{-} - \delta_{\phi, -}}{\zeta_{-}}.
\end{align}

Thus, consistency with De Giuli et al requires that 

\begin{equation}
\zeta_{-} = 2 \delta_{\phi, -}. \label{eq:twice}
\end{equation}

Note that $\zeta_+$ and $\delta_{\phi, +}$ are already known to satisfy the same requirement.

In the hard-sphere limit ($\Delta \phi < 0, T \to 0$), dimensional analysis requires that $P \propto T$.  Thus we extract the asymptotic equation of state

\begin{align}
P &\propto \left|\Delta \phi\right|^{\zeta_{-}/\delta_{\phi, -}- 1} \left(\frac{T}{\left|\Delta \phi\right|^{\zeta_{-}/ \delta_{\phi,-}}}\right) \propto \frac{T}{\left| \Delta \phi \right|}
\end{align}

Thus, the hard-sphere pressure scaling is automatically satisfied.

The pressure scaling in the regimes that De Giuli et al call `soft-zero-T`` and ``soft-entropic'' are obtained by taking $T\to 0$ for positive $\Delta \phi$; this recovers the previous zero-temperature scaling ansatz, giving $P \propto \Delta \phi$ in agreement with their results.

Now we must verify the scaling of the shear modulus $G$. At zero strain, it obeys the scaling form

\begin{align}
G &=\Delta Z^{\zeta_{\pm} - 2 \delta_{\epsilon, \pm}} \mathcal{G}_{\pm}{\left(\frac{\Delta \phi}{\Delta Z^\delta_{\phi, \pm}}, \frac{T}{\Delta Z^{\zeta_{\pm}}}\right)}.
\end{align}

The anharmonic and hard-sphere limits are treated identically to the case of the pressure, demanding a finite value at zero $\Delta \phi$ in the anharmonic case while requiring $G\propto T$ in the hard-sphere case.  Matching the anharmonic limit for $G$ with De Giuli et al turns out to require

\begin{equation}
\frac{\delta_{\epsilon, -}}{\zeta_{-}} = \frac{1}{4} \frac{4 + 2 \theta_e}{3 + \theta_e},
\end{equation}

while the hard-sphere limit requires

\begin{equation}
\frac{\delta_{\epsilon, -}}{\delta_{\phi, -}} = \frac{1}{2}\frac{4 + 2 \theta_e}{3 + \theta_e}.
\end{equation}

Note that these equations are consistent with equation \ref{eq:twice}, and are both satisfied by the exponents in Table \ref{tab:exp}.

The ``soft entropic'' regime of De Giuli et al. is less obvious, but is still consistent with the scaling ansatz in the $-$ branch. After eliminating $\Delta Z$ we have 

\begin{align}
G &= \left| \Delta \phi\right|^{2 - 2 \delta_{\epsilon, -} / \delta_{\phi, -}} \mathcal{G}_{-}{\left(\frac{T}{\Delta \phi^{2}}\right)} \\
&= T^{1 - 2 \delta_{\epsilon, -}/ \delta_{\phi, -}} \left| \Delta \phi\right|^{2 \delta_{\epsilon, -}/ \delta_{\phi, -}} g_{-}{\left(\frac{T}{\Delta \phi^{2}}\right)}.
\end{align}

If the function $g_{-}$ approaches a finite limit, this matches the ``soft entropic'' scaling as long as 

\begin{equation}
\frac{\delta_{\epsilon, -}}{\delta_{\phi, -}} = \frac{1}{2}\frac{4 + 2 \theta_e}{3 + \theta_e},
\end{equation}

which shows consistency with the exponents in the hard-sphere and anharmonic regimes.

To see the ``soft entropic'' scaling in our picture we must move along the phase boundary, $T = C^* \Delta \phi^2$, since our picture does not have the anomalous temperature scale $T^*$ of De Giuli et al;  $g_{-}$ is constant along this trajectory, as required. 

\section{Shear stress fluctuations at zero $\Delta \phi$}
\label{app:shear_stress}
At finite $T$, in the limit of small $\Delta \phi$ the scaling form for the shear stress fluctuations $s^2$ is

\begin{align}
s^2 &=\left|\Delta \phi \right|^{2 \left(\zeta_{-} -  \delta_{\epsilon, -}\right) / \delta_{\phi, -}} \mathcal{S}_{-}{\left(N \left| \Delta \phi \right|^{\psi_{-}/\delta_{\phi, -}}, \frac{T}{\left|\Delta \phi\right|^{2}}\right)}.
\end{align}

A finite correlation length requires $1/\sqrt{N}$ fluctuations of shear-stress for large $N$:

\begin{align}
s^2 &=\frac{1}{N}\left|\Delta \phi \right|^{\left(2\zeta_{-} -  2\delta_{\epsilon, -} -\psi_{-} \right) / \delta_{\phi, -}} \mathcal{\tilde{S}}_{-}{\left(\frac{T}{\left|\Delta \phi\right|^2}\right)}.
\end{align}

Finally, at finite $T$ the fluctuations cannot diverge or vanish as $\Delta \phi \to 0$. Imposing the required power law to cancel all factors of $\Delta \phi$ yields 

\begin{align}
s^2 N &\propto T^{\left(2 \zeta_{-} -  2\delta_{\epsilon, -} -\psi_{-} \right) /2 \delta_{\phi, -}}  = T^{\frac{7 + \theta_{e}}{6+2\theta_{e}}}
\end{align}

\section{The necessary form of the small-$u$ scaling function}
\label{app:u_exponents}
Note that

\begin{align}
    \zeta_{-} - \zeta_{+} &= \frac{2- 2\theta_{e}}{1+\theta_e} \\
    \delta_{\phi, -} - \delta_{\phi, +} &= \frac{1-\theta_{e}}{1+\theta_e} =  \frac{1}{2} \left( \zeta_{-} - \zeta_{+}\right) \\
    \delta_{\varepsilon, -} - \delta_{\varepsilon, +} &= \frac{1-\theta_{e}}{2+2\theta_e}  = \frac{1}{4} \left( \zeta_{-} - \zeta_{+}\right).
\end{align}

Thus, the $-$ exponents are recovered if the correction to scaling is, for small $u$, a function of $X_T/X_u$, $X_\phi/X_u^{1/2}$, and $X_\phi/X_u^{1/4}$.

\end{document}